\documentstyle[11pt,psfig,epic,eepic]{article}
\psfigurepath{/home/usuarios/fernando/PAPERS/TSALLIS/FIG}

\begin{document}

{\Large
{\bf Generalized Thermostatistical Description of Intermittency and} \\
{\bf Non-extensivity in Turbulence and Financial Markets} 
}
\bigskip
        
F. M. Ramos\footnote{e-mail:fernando@lac.inpe.br}, C. Rodrigues Neto and R. R. Rosa
\smallskip

{\small

Laborat\'orio Associado de Computa\c c\~ao e Matem\'atica Aplicada (LAC) \\
Instituto Nacional de Pesquisas Espaciais (INPE) \\
S\~ao Jos\'e dos Campos - SP, Brazil 

\bigskip

PACS.02.50-r - Probability theory, stochastic processes, and statistics. \\
PACS.47.27Eq - Turbulence simulation and modeling. \\
PACS.89.90+n - Other areas of general interest to physicists.

}
\bigskip


{\small{\bf Abstract.} - We present a new framework for modeling the statistical 
behavior of both fully developed turbulence and short-term dynamics of financial 
markets based on the generalized non-extensive thermostatistics formalism. 
We also show that intermittency -- strong bursts in the energy 
dissipation or clusters of high price volatility -- and non-extensivity -- 
anomalous scaling of usually extensive properties like entropy -- are
naturally linked by a single parameter $q$, from the
non-extensive thermostatistics.}

\bigskip

Scaling invariance plays a fundamental role in many natural 
phenomena and frequently emerges from some sort of underlying 
cascade process. A classical example is fully developed homogeneous 
isotropic three-dimensional turbulence, which is characterized by a 
cascade of kinetic energy from large forcing scales to smaller and 
smaller ones through a hierarchy of eddies. At the end
of the cascade, the energy dissipates by viscosity, turning into heat.
Recently, some authors [1-3] have studied the phenomenological
relationship among financial market dynamics, scaling behavior
and  hydrodynamic turbulence. Particularly, Ghashghaie {\it et al.} [2] 
conjectured the existence of a temporal information cascade similar 
to the spatial energy cascade found in fully developed turbulence.  

Traditionally, the properties of turbulent flows 
are studied from the statistics of velocity differences 
$v_r(x) = v(x) - v(x+r)$ at different scales $r$. As with other
physical systems that depend on the dynamical evolution of a large number
of nonlinearly coupled subsystems, the energy cascade in turbulence
generates a spatial scaling behavior -- power-law behavior with $r$ --
of the moments $\langle v_r^n \rangle$ of the probability distribution 
function (PDF) of $v_r$ (the angle brackets $\langle \rangle$ denote
the mean value of the enclosed quantity). For large values of the Reynolds number, 
which measures 
the ratio of nonlinear inertial forces to the linear dissipative 
forces within the fluid, there is a wide separation between the scale 
of energy input 
(integral scale $L$) and the viscous dissipation scale (Kolmogorov 
scale $\eta$). Though at large scales ($\sim L$) the PDFs are 
normally distributed, far from the integral scale they are strongly 
non-Gaussian and display wings fatter than expected for a normal process. 
This is the striking signature of the intermittency phenomenon. 
After publication of the Kolmogorov K62 refined similarity hypotheses [4],
the problem of small scale intermittency became one of the
central questions on isotropic turbulence. Over the past years 
several papers [5-12] have discussed intermittency and 
the so-called `PDF problem'. Similar 
attempts [1,2,13] have been made to explain the same peculiar shape observed in the  
PDF of price changes $z_{\tau} = z(t) - z(t + \tau)$ at small time intervals.

Based on the scaling properties of multifractals, a generalization of Boltzmann-Gibbs 
thermostatistics has been proposed [14-17] through the introduction of a family 
of non-extensive entropy functionals $S_q(p)$ with a single parameter $q$. These 
functionals reduce to the classical, 
extensive Boltzmann-Gibbs form as $q \rightarrow 1$. Optimizing $S_q[p]$ subject 
to appropriate
constraints [16], we obtain the distribution 
\begin{equation}
\label{pdf}
p_q(x) = [1 - \beta (1-q) x^2]^{1/(1-q)}/Z_q~~.
\end{equation}
The normalization factor, for $1 < q < 3$, is given by 
\[
Z_q \equiv \left[ \frac{\pi}{\beta (q-1)} \right]^{1/2} 
\frac{\Gamma((3-q)/2(q-1))}{\Gamma(1/(q-1))}~~. 
\]
In the limit of $q \rightarrow 1$, we recover the Gaussian distribution.  

The above distribution, we claim, provides a simple and accurate model
for handling the PDF problem. To show this, we stay in the context of fully 
developed turbulence  ($x \equiv v_r$). From equation (\ref{pdf}), we can  
easily obtain the second moment 
\begin{equation}
\label{sec}
\langle v_r^2 \rangle = \frac{1}{\beta (5 - 3 q)}~~,
\end{equation}
and the flatness coefficient
\begin{equation}
\label{kur}
K_r = \frac{\langle v_r^4 \rangle}{\langle v_r^2 \rangle^2} 
= \frac{3 \, (5 - 3 q)}{(7 -  5 q)}~~.
\end{equation}
We remark that the flatness coefficient, which is directly related to 
the occurrence of intermittency, is solely determined by the parameter $q$.

At this point, if we assume [2,5,6] a scaling of the moments 
$\langle v_r^n \rangle$ of $v_r$ as $r^{\zeta_n}$, 
the variation with $r$ of the PDF of the velocity 
differences and of its related moments can be completely determined. 
Particularly, we can obtain the functional forms of the flatness coefficient 
and the parameter $q$, respectively
\begin{equation}
\label{kur2}
K_r = K_L (r/L)^{\alpha}   
\end{equation} 
and
\begin{equation}
\label{qq}
q =\frac{15 - 7 K_L (r/L)^{\alpha}}{9 - 5 K_L (r/L)^{\alpha}} =
\frac{15 - 7 K_\eta (r/\eta)^{\alpha}}{9 - 5 K_\eta (r/\eta)^{\alpha}}~~,
\end{equation}
where $K_\eta$ is given by equation (\ref{kur2}), 
$\alpha = \zeta_4 - 2 \, \zeta_2$ 
and $K_L =3$, the expected value for 
a Gaussian process. The correspondent expression 
for $\beta$ can be derived similarly from equation (\ref{sec}). For a PDF 
normalized to its standard deviation (that is, with unit
variance), we have $\beta = 1/(5-3q)$.

We note that in the limit of infinite Reynolds numbers, as $r \rightarrow 0$, 
$K_r$ diverges (since $\alpha < 0$) while $q$ tends to a finite limit, 
$q < 7/5$. This bound coincides with the one obtained 
by Boghossian [18] through a $q$-generalization of 
Navier-Stokes equations. Moreover, this limit implies that the second moment
of distribution (\ref{pdf}) will always remain finite, which is empirically 
expected from the phenomena here analyzed.
 
In order to account for the well known asymmetry of the 
velocity distributions, we may also consider $\beta=\beta_p$, for $v_r \geq 0$, 
and $\beta=\beta_m$, for $v_r < 0$. In this case, we have 
\begin{equation}
\label{asim}
\langle v_r^n \rangle = A_n(\beta_m,\beta_p) \, s^{n/2} \, 
\frac{\Gamma(\frac{n+1}{2})\Gamma(s - \frac{n+1}{2})}
{\Gamma(\frac{1}{2})\Gamma(s - \frac{1}{2})}~~,
\end{equation}
where $s=1/(q-1)$ and 
\begin{equation}
A_n(\beta_m,\beta_p)=\frac{\beta_p^{-(n+1)/2} + (-1)^n \, \beta_m^{-(n+1)/2}}
{\beta_p^{-1/2} + \beta_m^{-1/2}}~~.
\end{equation}
Accordingly, the value of parameters $q$, $\beta_m$ and $\beta_p$, at a given 
scale $r$, are now determined from the second moment, the skewness 
($S_r = \langle v_r^3 \rangle/\langle v_r^2 \rangle^{3/2}$) and 
the flatness coefficients. 

We checked our model with turbulence statistics data taken from 
reference [2], provided by Chabaud {\it et al.} [9]. 
For this, we compared the experimental data with the theoretical predictions, 
for an skewness factor at the integral scale of $S_L=-0.4$, $L=1$cm [9] and 
$\alpha=-0.10$ [5,19]. The results 
are displayed in Fig. 1a. A good agreement is observed through spatial scales
spanning two orders of magnitude --
from the neighborhood of the integral scale down to close to the Kolmogorov 
scale --, and for a range of up to 15 standard deviations, including the rare fluctuations
in the tails of the distributions. Note that the solid lines in Fig. 1a have not 
been adjusted to the data through a free parameter, as in other models [9,6,7]. 
In the present case, the parameters $q$, $\beta_m$ and $\beta_p$, that 
control the shape of the PDF in each scale, are uniquely determined from the scaling 
of $\langle v_r^2 \rangle$, $S_r$ and $K_r$, obtained from equation (\ref{asim}).

The same approach adopted in turbulence can be 
straightforwardly applied (with $x \equiv z_{\tau}$ and $z_{\tau}$ scaling as 
$\tau^{\xi_n}$) 
to model the statistics of price differences in 
financial markets, as far as the parameter $\alpha$, $S_L$ and the integral time 
scale $\tau_L$ -- time span for which a convergence to a Gaussian process is found --
are available. We tested our model with price changes data taken from 
reference [2], provided by Olsen \& Associates. The results are displayed 
in Fig. 1b, for $\alpha = -0.16$, $S_L=-0.4$ and $\tau_L \simeq 2.3$ days. 
Again, we observe that the proposed model reproduces with good
accuracy the statistics of price differences over all temporal scales. 

Non-extensivity, a matter of speculation in some areas [20], is an essential 
feature of the generalized thermostatistics. If we suppose a
scenario of a cascade of bifurcations with $m$ levels, and scale 
the generalized expectation value of an observable $O_q$ (the kinetic energy
$\frac{1}{2}v_r^2$ of velocity differences, for example), averaged over a 
volume of size $V=\eta^3$ and normalized by Boltzmann constant, we have at 
the first level [21]
\begin{equation}
O_q(2V) = 2 O_q(V) + 2 (1-q_0) O_q(V) S_q(V)
\end{equation}
and at level $m$
\[ O_q(2^m V) = 2 O_q(2^{m-1}V) + 2 (1-q_{m-1}) O_q(2^{m-1}V) S_q(2^{m-1}V) \]
\begin{equation}
\label{casca1}
\simeq 2^m O_q(V) + (1-q_{m-1}) 2^{m+C} O_q(V)~~,
\end{equation}
where the higher order terms in $(1-q)$ have been neglected, with $C$ being a constant
to be determined later. Cascade processes
are also described in terms of fractal or multifractals models [7,22-25]. 
Within these frameworks, in high Reynolds number turbulence, the energy
dissipation is not uniformly distributed within the fluid but rather
concentrated on subsets of non-integer fractal $D_F$ dimension. This picture 
leads to a scaling behavior 
with dimensionality not equal to the dimension $D$ of the embedding space. 
In this case, if we consider the cascade of bifurcations described above, we find
\begin{equation}
\label{casca2}
O_q(2^m V) = 2^{m D_F/D} O_q(V)~~,
\end{equation}
with $D=3$. It follows immediately from equations (\ref{casca1}) and (\ref{casca2}) 
that
\begin{equation}
D_F \simeq \frac{D}{m} \left[ \frac{\log(2^{-C} + 1 - q)}{\log(2)} + m + C \right]~~.
\end{equation}
At the top of the cascade, we have $q_{m-1}=1$
and $D_F = D$. $C$ is determined from the value of $D_F$ at the bottom of the cascade. 
Writing $2^{m-1} \eta \equiv r$ and $q_{m-1} \equiv 
q(r)$, we get
\begin{equation}
\label{casca3}
D_F \simeq D \, \frac{\log((2^{-C} + 1 - q(r))\, 2^{C+1} \, r/\eta)}{\log(2 \, r/\eta)}~~,
\end{equation}
where $q(r)$ is given by equation (\ref{qq}). 
For $C=-1$, at the bottom of the cascade ($r = \eta$), using the values 
of $\alpha$ and $L$ 
previously specified, and $\eta = 0.022$mm [9], we get $D_F \simeq 2.37$.
This value is in good agreement with the fractal dimension of interfaces
in turbulent flows ($D_F = 2.35 \pm 0.05$), measured in different 
experimental contexts [26-29].

Equation (\ref{casca3}), through the variation of parameter $q$, offers a 
quantitative picture of the transition from small-scale intermittent, 
non-extensive, fractal behavior to large-scale 
Gaussian, extensive homogeneity. This equation can also be exploited
to estimate the structure functions exponents. 
Under the assumptions of the multifractal model [30], $\zeta_n$ is 
given by $\zeta_n = \mbox{inf}_h [ n \, h + D - D_F(h) ]$, 
with $\zeta_3 = 1$, assuming that the turbulent flow possesses 
a range of scaling exponents $I = (h_{min},h_{max})$. 
The existence of a corresponding range of dissipation scales
$\eta^{\prime}$ extending from $\eta_{min} \sim L \, R^{-1/(1+h_{min})}$ to
$\eta_{max} \sim L \, R^{-1/(1+h_{max})}$, where $R$ is the Reynolds number
at the integral scale, allows us to rewrite   
\begin{equation}
\label{mf2}
\zeta_n = \begin{array}[t]{c}
          \mbox{inf} \\ {\footnotesize{\eta^{\prime}}} \end{array} 
          [ n \, h(\eta^{\prime}) + D - D_F(\eta^{\prime}) ] ~~,
\end{equation}
where $D_F(\eta^{\prime})$ is given by equation (\ref{casca3}), with 
$r \equiv \eta^{\prime}$. The resulting exponents $\zeta_n$ are in
good agreement with experimental values [19,5],
as shown in Table 1, for $n=2,\ldots,8$.

The above picture may also be applied to the 
information cascade, with $D=1$. One main qualitative difference 
between the two processes is that, since there is nothing equivalent
to viscous damping in the dynamics of speculative markets, the information 
cascade depth is only limited by the minimum time necessary to perform a 
trading transaction (roughly 1 min [1]). On the other hand, as in turbulence, 
the scaling exponents 
depend on the order of the moments in a nonlinear way. Figure 2 displays the 
comparison of the experimentally measured scaling exponents $\xi_n$ [2] and 
the theoretical prediction. We found a good coincidence between our results 
and the experimental data, assuming an equivalent Reynolds
number $R$ of approximately 9000 and $\xi_2 = 1.06$ [3].  

Summarizing, we described a simple and accurate framework for modeling 
the statistical behavior of both fully developed turbulence and short-term 
dynamics of financial 
markets based on the formalism of the generalized non-extensive thermostatistics. 
Within this framework, we have shown that intermittency and non-extensivity are
naturally linked by parameter $q$, which represents an objective measure 
of small-scale intermittent, fractal behavior in turbulent cascades.
 
\bigskip

{\bf References}

1. Mantegna, R.N. \& Stanley, H.E. Scaling behaviour in the dynamics of an
economic index. {\it Nature} {\bf 376}, 46-49 (1995).

2. Ghashghaie, S., Breymann, W., Peinke, J., Talkner, P. \& Dodge, Y. 
Turbulent cascades in foreign exchange markets. {\it Nature} {\bf 381}, 
767-770 (1996).

3. Mantegna, R.N. \& Stanley, H.E. Turbulence and financial markets. 
{\it Nature} {\bf 383}, 587-588 (1996).

4. Kolmogorov, A. M. A refinement of previous hypotheses concerning the 
local structure of turbulence of a viscous incompressible fluid at
high Reynolds number. {\it J. Fluid Mech.} {\bf 13}, 82-85 (1962).

5. Anselmet, F., Gagne, Y., Hopfinger, E. J. \& Antonia, R. A. High-order 
velocity structure functions in turbulent shear flows. {\it J. Fluid Mech.} 
{\bf 140}, 63-89 (1984).

6. Castaing, B., Gagne, Y. \& Hopfinger, E. J. Velocity probability 
density functions of high Reynolds number turbulence. {\it Physica D} 
{\bf 46}, 177-200 (1990).

7. Benzi, R., Biferale, L., Paladin, G., Vulpiani, A. \& Vergassola, M.
Multifractality in the statistics of the velocity gradients in turbulence.
{\it Phys. Rev. Lett.} {\bf 67}, 2299-2302 (1991).

8. Sreenivasan, K. R. \& Kailasnath, P. An update on the intermittency
exponent in turbulence. {\it Phys. Fluids} {\bf 5}, 512-514 (1992).
  
9. Chabaud et al. Transition towards developed turbulence. 
{\it Phys. Rev. Lett.} {\bf 73}, 3227-3230, (1994). 

10. Praskovsky, A. \& Oncley, S. Measurements of the Kolmogorov constant and
intermittency exponent at very high Reynolds number. {\it Phys. Fluids}
{\bf 6}, 2886-2888 (1994)

11. Vassilicos, J.C. Turbulence and intermittency 
{\it Nature} {\bf 374}, 408-409 (1995).

12. Naert, A., Castaing, B., Chabaud, B., B. H\'ebral \& Peinke, J.
Conditional statistics of velocity fluctuations in turbulence. {\it Physica D} 
{\bf 113}, 73-78 (1998).

13. Taylor, S. J. {\it Math. Fin.} Modeling stochastic volatility: a review
and comparative study. {\bf 4}, 183-204 (1994).

14. Tsallis, C. Possible generalization of Boltzmann-Gibbs statistics. 
{\it J. Stat. Phys.} {\bf 52}, 479-487 (1988).

15. Tsallis, C., S\'a Barreto, F.C. \& Loh, E.D. Generalization of the Planck
radiation law and application to the cosmic microwave background radiation. 
{\it Phys. Rev. E} {\bf 52}, 1447-1451 (1995).

16. Tsallis, C., Levy, S. V. F., Souza, A. M. C. \& Maynard, R. 
Statistical-mechanical foundation of the ubiquity of L\'evy 
distributions in nature. {\it Phys. Rev. Lett.} {\bf 75}, 3589-3593 (1995).

17. Lyra, M. L. \& Tsallis, C. Non-extensivity and multifractality in 
low-dimensional dissipative systems {\it Phys. Rev. Lett.} {\bf 80}, 53-56 
(1998).

18. Boghossian, B. M., Navier-Stokes equations for generalized thermostatistics.
{\it Braz. J. Phys.} {\bf 29}, 91-107 (1999).

19. Benzi, R., Ciliberto, S., Baudet, C. \& and Chavarria, C. R. On the
scaling of three-dimensional homogeneous and isotropic turbulence. 
{\it Physica D} {\bf 80}, 385-398 (1995).

20. Maddox, J., When entropy does not seem extensive. {\it Nature} 
{\bf 365}, 103- (1993).

21. Boghossian, B. M., Thermodynamic description of the relaxation of 
two-dimensional Euler turbulence using Tsallis statistics. {\it Phys. Rev. E} 
{\bf 53}, 4754 (1996).

22. Mandelbrot, B. B. Intermittent turbulence in self-similar cascades: 
divergence of high moments and dimension of the carrier. {\it J. Fluid Mech.} 
{\bf 62}, 331-358 (1974).

23. Parisi, G. \& Frisch, U. in {\it Turbulence and Predictability in
Geophysical Fluid Dynamics and Climatic Dynamics} (ed. Ghil, M., Benzi, R.,
\& Parisi, G.) 84- (North-Holland, Amsterdam, 1985).
  
24. Paladin, G. \& Vulpiani, A. Anomalous scaling laws in multifractal objects.
{\it Phys. Rep.} {\bf 156}, 147-225 (1987).

25. Meneveau, C. \& Sreenivasan, K. R. The multifractal nature of turbulent 
energy dissipation. {\it J. Fluid Mech.} {\bf 224}, 
429-484 (1991).

26. Hentschel, H. G. E. \& Procaccia, I. Relative diffusion in turbulent media:
the fractal dimension of clouds. {\it Phys. Rev. A} {\bf 29}, 1461-1470, (1984). 

27. Sreenivasan, K. R. \& Meneveau, C. Fractal facets of turbulence. 
{\it J. Fluid Mech.} {\bf 173}, 357-386 (1986).
 
28. Sreenivasan, K. R., Ramshankar, R.  \& Meneveau, C. Mixing, entrainment and 
fractal dimensions of surfaces in turbulent flows. {\it Proc. R. Soc. Lond. A}
{\bf 421}, 79-108 (1989).

29. Vassilicos, J. C. \& Hunt, J. C. R. Fractal dimensions and spectra of 
interfaces with application to turbulence. {\it Proc. R. Soc. Lond. A}
{\bf 435}, 505-534 (1991).

30. Frisch, U. {\it Turbulence} (Cambridge Univ. Press, Cambridge, UK, 1995).

\bigskip

{\bf Acknowledgments:} We thank C. Tsallis for helpful discussions. This
work was supported by FAPESP-Brazil and CNPq-Brazil.

\bigskip

\newpage

{\bf Captions}

{\bf Table 1} Comparison of experimentally measured scaling exponents $\zeta_n$ 
and the theoretical prediction.

\bigskip

{\bf Figure 1(a)} Data points: standardized probability distributions $p_q(v_r)$
of velocity differences $v_r(x) = v(x) - v(x+r)$ for spatial scales, from top
to bottom, $r/\eta = 3.3$, $18.5$, $138$ and $325$ (data taken from ref. [2], 
provided by Chabaud {\it et al.} [9]); Solid lines: proposed PDF model, equation 
(\ref{pdf}) with $\beta=\beta_p$, for $v_r \geq 0$, and $\beta=\beta_m$, 
for $v_r < 0$  (for better visibility the curves have 
been vertically shifted with respect to each other).

\bigskip

{\bf Figure 1(b)} Data points: standardized probability distributions 
$p_q(z_{\tau})$ of price differences $z_{\tau}(t) = z(t) - z(t+\tau)$ 
for temporal scales, from top to bottom, $\tau = 640~s$, $5120~s$, $40960~s$ 
and $163840~s$ (data taken from ref. [2], provided by Olsen \& Associates); 
Solid lines: proposed PDF model, equation 
(\ref{pdf}) with $\beta=\beta_p$, for $z_{\tau} \geq 0$, and $\beta=\beta_m$, 
for $z_{\tau} < 0$  (for better visibility the curves have 
been vertically shifted with respect to each other).

\bigskip

{\bf Figure 2} Comparison of theoretical and experimentally measured scaling 
exponents $\xi_n$.

\newpage

\begin{table}[hhh]
\begin{center}
\begin{tabular}{cccc}  \hline
 Order & Theory & Experiment  & Experiment \\ 
 $n$ & $\zeta_n$ & [19] & [5] \\ \hline
 2 & 0.72 & 0.70 & 0.71 \\
 3 & 1.00 & 1.00 & 1.00 \\
 4 & 1.27 & 1.28 & 1.33 \\
 5 & 1.54 & 1.54 & 1.65 \\
 6 & 1.81 & 1.78 & 1.80 \\
 7 & 2.07 & 2.00 & 2.12 \\
 8 & 2.33 & 2.23 & 2.22 \\ 
\hline
\end{tabular}
\end{center}
\end{table}

\newpage

\begin{figure}[ppp]
\begin{center}
\input{fig1a}
\end{center}
\end{figure}

\newpage

\begin{figure}[ppp]
\begin{center}
\input{fig1b}
\end{center}
\end{figure}

\newpage

\begin{figure}[ppp]
\begin{center}
\setlength{\unitlength}{0.240900pt}
\ifx\plotpoint\undefined\newsavebox{\plotpoint}\fi
\sbox{\plotpoint}{\rule[-0.200pt]{0.400pt}{0.400pt}}%
\begin{picture}(1800,1350)(0,0)
\font\gnuplot=cmr10 at 10pt
\gnuplot
\sbox{\plotpoint}{\rule[-0.200pt]{0.400pt}{0.400pt}}%
\put(220.0,113.0){\rule[-0.200pt]{365.204pt}{0.400pt}}
\put(220.0,113.0){\rule[-0.200pt]{0.400pt}{292.453pt}}
\put(220.0,113.0){\rule[-0.200pt]{4.818pt}{0.400pt}}
\put(198,113){\makebox(0,0)[r]{$0$}}
\put(1716.0,113.0){\rule[-0.200pt]{4.818pt}{0.400pt}}
\put(220.0,265.0){\rule[-0.200pt]{4.818pt}{0.400pt}}
\put(198,265){\makebox(0,0)[r]{$0.5$}}
\put(1716.0,265.0){\rule[-0.200pt]{4.818pt}{0.400pt}}
\put(220.0,417.0){\rule[-0.200pt]{4.818pt}{0.400pt}}
\put(198,417){\makebox(0,0)[r]{$1$}}
\put(1716.0,417.0){\rule[-0.200pt]{4.818pt}{0.400pt}}
\put(220.0,568.0){\rule[-0.200pt]{4.818pt}{0.400pt}}
\put(198,568){\makebox(0,0)[r]{$1.5$}}
\put(1716.0,568.0){\rule[-0.200pt]{4.818pt}{0.400pt}}
\put(220.0,720.0){\rule[-0.200pt]{4.818pt}{0.400pt}}
\put(198,720){\makebox(0,0)[r]{$2$}}
\put(1716.0,720.0){\rule[-0.200pt]{4.818pt}{0.400pt}}
\put(220.0,872.0){\rule[-0.200pt]{4.818pt}{0.400pt}}
\put(198,872){\makebox(0,0)[r]{$2.5$}}
\put(1716.0,872.0){\rule[-0.200pt]{4.818pt}{0.400pt}}
\put(220.0,1024.0){\rule[-0.200pt]{4.818pt}{0.400pt}}
\put(198,1024){\makebox(0,0)[r]{$3$}}
\put(1716.0,1024.0){\rule[-0.200pt]{4.818pt}{0.400pt}}
\put(220.0,1175.0){\rule[-0.200pt]{4.818pt}{0.400pt}}
\put(198,1175){\makebox(0,0)[r]{$3.5$}}
\put(1716.0,1175.0){\rule[-0.200pt]{4.818pt}{0.400pt}}
\put(220.0,1327.0){\rule[-0.200pt]{4.818pt}{0.400pt}}
\put(198,1327){\makebox(0,0)[r]{$4$}}
\put(1716.0,1327.0){\rule[-0.200pt]{4.818pt}{0.400pt}}
\put(220.0,113.0){\rule[-0.200pt]{0.400pt}{4.818pt}}
\put(220,68){\makebox(0,0){$0$}}
\put(220.0,1307.0){\rule[-0.200pt]{0.400pt}{4.818pt}}
\put(523.0,113.0){\rule[-0.200pt]{0.400pt}{4.818pt}}
\put(523,68){\makebox(0,0){$2$}}
\put(523.0,1307.0){\rule[-0.200pt]{0.400pt}{4.818pt}}
\put(826.0,113.0){\rule[-0.200pt]{0.400pt}{4.818pt}}
\put(826,68){\makebox(0,0){$4$}}
\put(826.0,1307.0){\rule[-0.200pt]{0.400pt}{4.818pt}}
\put(1130.0,113.0){\rule[-0.200pt]{0.400pt}{4.818pt}}
\put(1130,68){\makebox(0,0){$6$}}
\put(1130.0,1307.0){\rule[-0.200pt]{0.400pt}{4.818pt}}
\put(1433.0,113.0){\rule[-0.200pt]{0.400pt}{4.818pt}}
\put(1433,68){\makebox(0,0){$8$}}
\put(1433.0,1307.0){\rule[-0.200pt]{0.400pt}{4.818pt}}
\put(1736.0,113.0){\rule[-0.200pt]{0.400pt}{4.818pt}}
\put(1736,68){\makebox(0,0){$10$}}
\put(1736.0,1307.0){\rule[-0.200pt]{0.400pt}{4.818pt}}
\put(220.0,113.0){\rule[-0.200pt]{365.204pt}{0.400pt}}
\put(1736.0,113.0){\rule[-0.200pt]{0.400pt}{292.453pt}}
\put(220.0,1327.0){\rule[-0.200pt]{365.204pt}{0.400pt}}
\put(45,720){\makebox(0,0){$\xi_n$}}
\put(978,23){\makebox(0,0){$n$}}
\put(220.0,113.0){\rule[-0.200pt]{0.400pt}{292.453pt}}
\put(1606,1262){\makebox(0,0)[r]{Present model}}
\put(1650,1262){\raisebox{-.8pt}{\makebox(0,0){$\Diamond$}}}
\put(523,435){\raisebox{-.8pt}{\makebox(0,0){$\Diamond$}}}
\put(675,517){\raisebox{-.8pt}{\makebox(0,0){$\Diamond$}}}
\put(826,589){\raisebox{-.8pt}{\makebox(0,0){$\Diamond$}}}
\put(978,656){\raisebox{-.8pt}{\makebox(0,0){$\Diamond$}}}
\put(1130,718){\raisebox{-.8pt}{\makebox(0,0){$\Diamond$}}}
\put(1281,777){\raisebox{-.8pt}{\makebox(0,0){$\Diamond$}}}
\put(1433,833){\raisebox{-.8pt}{\makebox(0,0){$\Diamond$}}}
\put(1584,887){\raisebox{-.8pt}{\makebox(0,0){$\Diamond$}}}
\put(1736,939){\raisebox{-.8pt}{\makebox(0,0){$\Diamond$}}}
\sbox{\plotpoint}{\rule[-0.400pt]{0.800pt}{0.800pt}}%
\put(1606,1217){\makebox(0,0)[r]{Experiment [2]}}
\put(1628.0,1217.0){\rule[-0.400pt]{15.899pt}{0.800pt}}
\put(523,365){\usebox{\plotpoint}}
\multiput(524.41,365.00)(0.501,0.549){297}{\rule{0.121pt}{1.079pt}}
\multiput(521.34,365.00)(152.000,164.761){2}{\rule{0.800pt}{0.539pt}}
\multiput(675.00,533.41)(1.819,0.502){77}{\rule{3.076pt}{0.121pt}}
\multiput(675.00,530.34)(144.615,42.000){2}{\rule{1.538pt}{0.800pt}}
\multiput(826.00,575.41)(1.140,0.501){127}{\rule{2.015pt}{0.121pt}}
\multiput(826.00,572.34)(147.818,67.000){2}{\rule{1.007pt}{0.800pt}}
\multiput(978.00,642.41)(0.761,0.501){193}{\rule{1.416pt}{0.121pt}}
\multiput(978.00,639.34)(149.061,100.000){2}{\rule{0.708pt}{0.800pt}}
\multiput(1130.00,742.41)(1.186,0.501){121}{\rule{2.088pt}{0.121pt}}
\multiput(1130.00,739.34)(146.667,64.000){2}{\rule{1.044pt}{0.800pt}}
\multiput(1281.00,806.41)(2.143,0.503){65}{\rule{3.578pt}{0.121pt}}
\multiput(1281.00,803.34)(144.574,36.000){2}{\rule{1.789pt}{0.800pt}}
\multiput(1433.00,842.41)(2.070,0.503){67}{\rule{3.465pt}{0.121pt}}
\multiput(1433.00,839.34)(143.809,37.000){2}{\rule{1.732pt}{0.800pt}}
\multiput(1584.00,879.41)(0.930,0.501){157}{\rule{1.683pt}{0.121pt}}
\multiput(1584.00,876.34)(148.507,82.000){2}{\rule{0.841pt}{0.800pt}}
\put(1650,1217){\raisebox{-.8pt}{\makebox(0,0){$\Box$}}}
\put(523,365){\raisebox{-.8pt}{\makebox(0,0){$\Box$}}}
\put(675,532){\raisebox{-.8pt}{\makebox(0,0){$\Box$}}}
\put(826,574){\raisebox{-.8pt}{\makebox(0,0){$\Box$}}}
\put(978,641){\raisebox{-.8pt}{\makebox(0,0){$\Box$}}}
\put(1130,741){\raisebox{-.8pt}{\makebox(0,0){$\Box$}}}
\put(1281,805){\raisebox{-.8pt}{\makebox(0,0){$\Box$}}}
\put(1433,841){\raisebox{-.8pt}{\makebox(0,0){$\Box$}}}
\put(1584,878){\raisebox{-.8pt}{\makebox(0,0){$\Box$}}}
\put(1736,960){\raisebox{-.8pt}{\makebox(0,0){$\Box$}}}
\sbox{\plotpoint}{\rule[-0.500pt]{1.000pt}{1.000pt}}%
\put(1606,1172){\makebox(0,0)[r]{n/3}}
\multiput(1628,1172)(20.756,0.000){4}{\usebox{\plotpoint}}
\put(1694,1172){\usebox{\plotpoint}}
\put(220,113){\usebox{\plotpoint}}
\put(220.00,113.00){\usebox{\plotpoint}}
\put(237.31,124.45){\usebox{\plotpoint}}
\put(254.72,135.73){\usebox{\plotpoint}}
\put(271.63,147.76){\usebox{\plotpoint}}
\put(289.05,159.03){\usebox{\plotpoint}}
\put(306.47,170.32){\usebox{\plotpoint}}
\put(323.38,182.35){\usebox{\plotpoint}}
\put(340.79,193.62){\usebox{\plotpoint}}
\multiput(343,195)(17.270,11.513){0}{\usebox{\plotpoint}}
\put(358.11,205.07){\usebox{\plotpoint}}
\put(375.38,216.58){\usebox{\plotpoint}}
\put(392.60,228.16){\usebox{\plotpoint}}
\put(409.76,239.84){\usebox{\plotpoint}}
\put(427.03,251.35){\usebox{\plotpoint}}
\put(444.50,262.56){\usebox{\plotpoint}}
\put(461.50,274.44){\usebox{\plotpoint}}
\put(478.66,286.11){\usebox{\plotpoint}}
\multiput(480,287)(17.601,11.000){0}{\usebox{\plotpoint}}
\put(496.23,297.15){\usebox{\plotpoint}}
\put(513.50,308.67){\usebox{\plotpoint}}
\put(530.73,320.25){\usebox{\plotpoint}}
\put(547.89,331.92){\usebox{\plotpoint}}
\put(565.15,343.44){\usebox{\plotpoint}}
\put(582.62,354.64){\usebox{\plotpoint}}
\put(599.99,366.00){\usebox{\plotpoint}}
\put(616.82,378.14){\usebox{\plotpoint}}
\multiput(618,379)(17.270,11.513){0}{\usebox{\plotpoint}}
\put(634.08,389.67){\usebox{\plotpoint}}
\put(651.63,400.75){\usebox{\plotpoint}}
\put(668.75,412.48){\usebox{\plotpoint}}
\put(685.82,424.26){\usebox{\plotpoint}}
\put(703.26,435.51){\usebox{\plotpoint}}
\put(720.53,447.02){\usebox{\plotpoint}}
\put(738.05,458.15){\usebox{\plotpoint}}
\put(754.93,470.21){\usebox{\plotpoint}}
\multiput(756,471)(17.270,11.513){0}{\usebox{\plotpoint}}
\put(772.19,481.74){\usebox{\plotpoint}}
\put(789.73,492.82){\usebox{\plotpoint}}
\put(806.85,504.56){\usebox{\plotpoint}}
\put(823.93,516.33){\usebox{\plotpoint}}
\put(841.37,527.58){\usebox{\plotpoint}}
\put(858.64,539.09){\usebox{\plotpoint}}
\put(875.91,550.60){\usebox{\plotpoint}}
\put(893.03,562.33){\usebox{\plotpoint}}
\multiput(894,563)(17.270,11.513){0}{\usebox{\plotpoint}}
\put(910.29,573.86){\usebox{\plotpoint}}
\put(927.63,585.27){\usebox{\plotpoint}}
\put(944.97,596.65){\usebox{\plotpoint}}
\put(961.92,608.61){\usebox{\plotpoint}}
\put(979.37,619.85){\usebox{\plotpoint}}
\put(996.76,631.17){\usebox{\plotpoint}}
\put(1014.03,642.69){\usebox{\plotpoint}}
\put(1031.15,654.42){\usebox{\plotpoint}}
\multiput(1032,655)(17.270,11.513){0}{\usebox{\plotpoint}}
\put(1048.42,665.94){\usebox{\plotpoint}}
\put(1065.76,677.35){\usebox{\plotpoint}}
\put(1083.26,688.50){\usebox{\plotpoint}}
\put(1100.29,700.35){\usebox{\plotpoint}}
\put(1117.32,712.21){\usebox{\plotpoint}}
\put(1134.81,723.38){\usebox{\plotpoint}}
\put(1152.16,734.77){\usebox{\plotpoint}}
\put(1168.95,746.96){\usebox{\plotpoint}}
\multiput(1169,747)(17.601,11.000){0}{\usebox{\plotpoint}}
\put(1186.52,758.01){\usebox{\plotpoint}}
\put(1203.79,769.53){\usebox{\plotpoint}}
\put(1221.18,780.86){\usebox{\plotpoint}}
\put(1238.40,792.42){\usebox{\plotpoint}}
\put(1255.42,804.28){\usebox{\plotpoint}}
\put(1272.92,815.45){\usebox{\plotpoint}}
\put(1290.26,826.84){\usebox{\plotpoint}}
\multiput(1292,828)(16.737,12.274){0}{\usebox{\plotpoint}}
\put(1307.06,839.04){\usebox{\plotpoint}}
\put(1324.63,850.08){\usebox{\plotpoint}}
\put(1341.90,861.60){\usebox{\plotpoint}}
\put(1359.17,873.11){\usebox{\plotpoint}}
\put(1376.35,884.74){\usebox{\plotpoint}}
\put(1393.55,896.37){\usebox{\plotpoint}}
\put(1410.82,907.88){\usebox{\plotpoint}}
\put(1428.36,918.97){\usebox{\plotpoint}}
\multiput(1430,920)(16.737,12.274){0}{\usebox{\plotpoint}}
\put(1445.18,931.12){\usebox{\plotpoint}}
\put(1462.50,942.56){\usebox{\plotpoint}}
\put(1480.02,953.68){\usebox{\plotpoint}}
\put(1497.29,965.19){\usebox{\plotpoint}}
\put(1514.48,976.83){\usebox{\plotpoint}}
\put(1531.68,988.45){\usebox{\plotpoint}}
\put(1548.95,999.96){\usebox{\plotpoint}}
\put(1566.49,1011.05){\usebox{\plotpoint}}
\multiput(1568,1012)(17.270,11.513){0}{\usebox{\plotpoint}}
\put(1583.76,1022.56){\usebox{\plotpoint}}
\put(1600.58,1034.72){\usebox{\plotpoint}}
\put(1617.94,1046.09){\usebox{\plotpoint}}
\put(1635.42,1057.28){\usebox{\plotpoint}}
\put(1652.42,1069.17){\usebox{\plotpoint}}
\put(1669.68,1080.68){\usebox{\plotpoint}}
\put(1687.05,1092.03){\usebox{\plotpoint}}
\put(1704.32,1103.55){\usebox{\plotpoint}}
\multiput(1705,1104)(17.601,11.000){0}{\usebox{\plotpoint}}
\put(1721.86,1114.63){\usebox{\plotpoint}}
\put(1736,1125){\usebox{\plotpoint}}
\end{picture}
\end{center}
\end{figure}

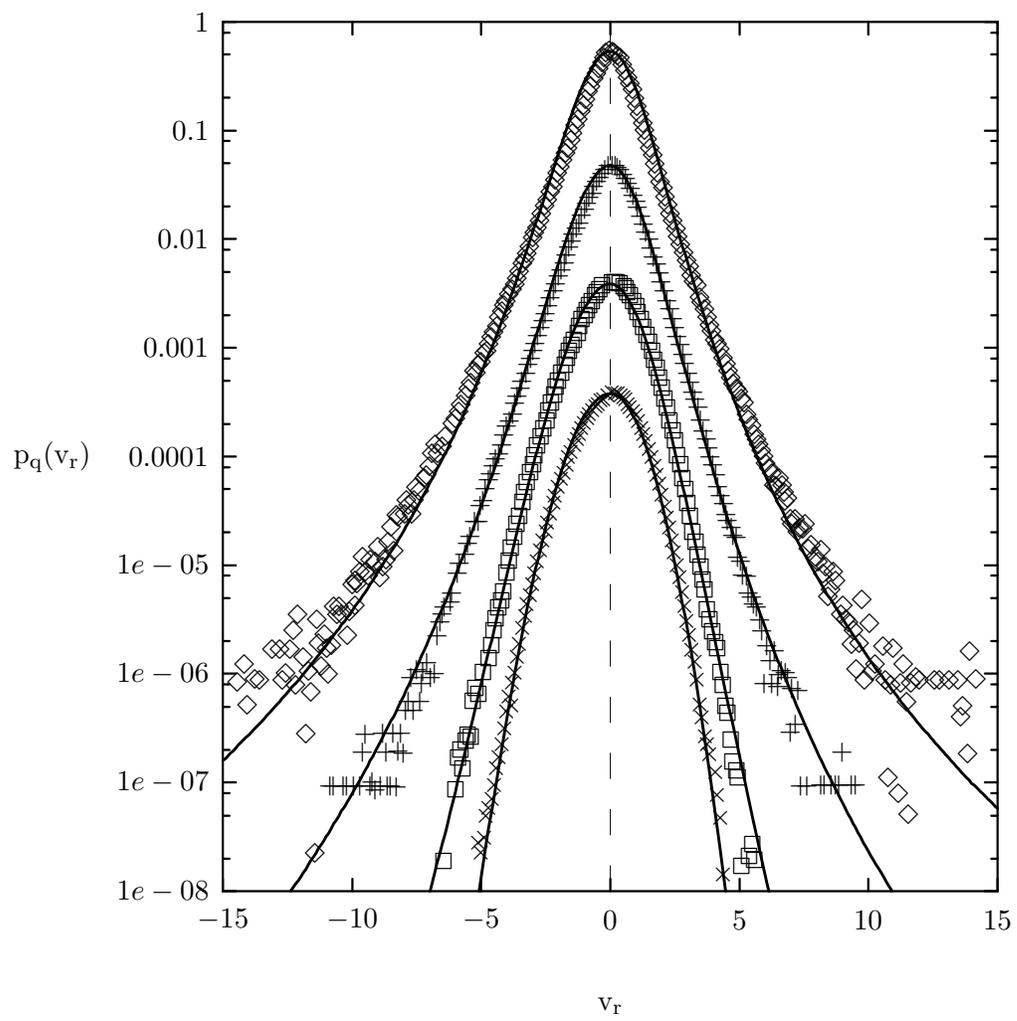
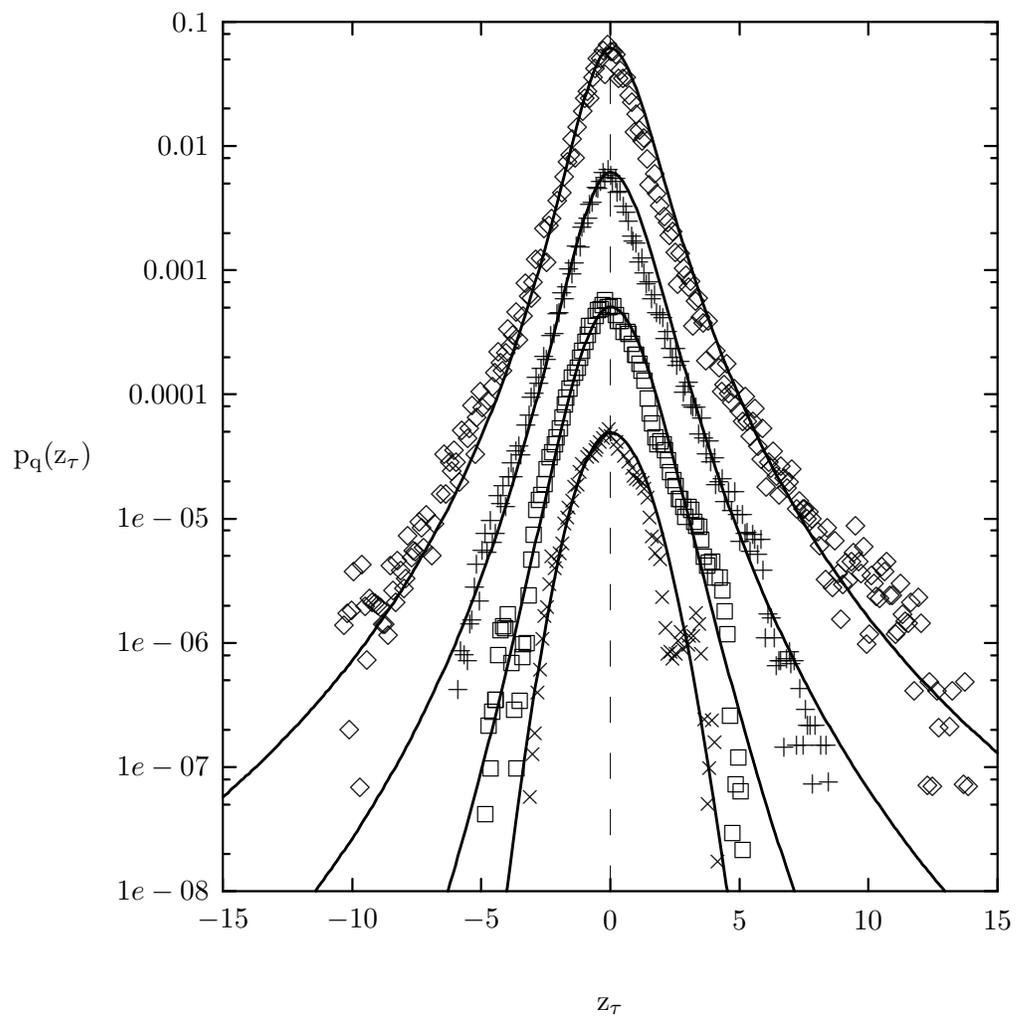
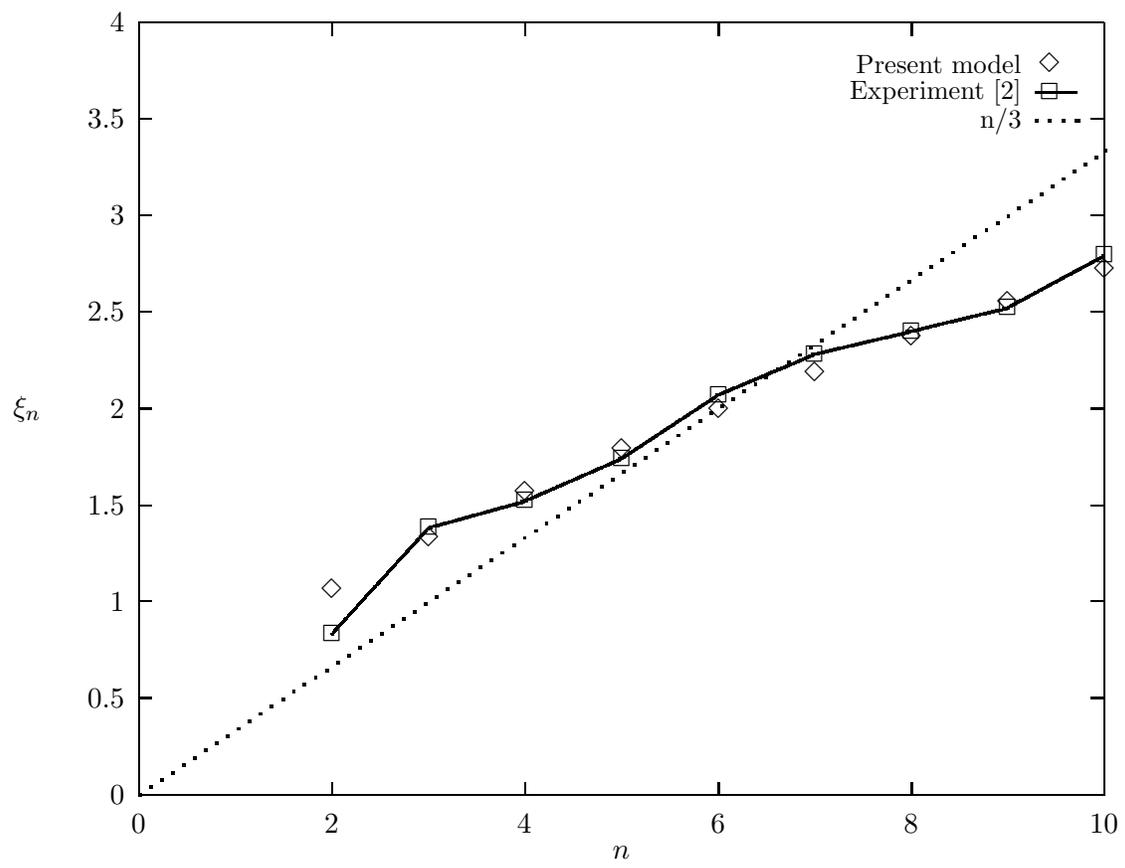
\end{document}